# The use of external controls: To what extent can it currently be recommended?[*]

Hans Ulrich Burger [1), Christoph Gerlinger [2)], Chris Harbron [3)], Armin Koch [4)], Martin Posch [5)], Justine Rochon [6)], Anja Schiel [7)]

[1)] Hoffmann-La Roche AG, [2)] Statistics and Data Insights, Bayer AG and Gynecology, Obstetrics and Reproductive Medicine, University Medical School of Saarland , [3)] Roche Products, [4)] Medizinische Hochschule Hannover, [5)] Center for Medical Statistics, Informatics, and Intelligent Systems, Medical University of Vienna, [6)] Boehringer Ingelheim Pharma GmbH & Co. KG, [7)] Norwegian Medicine Agency

**Abstract**

With more and better clinical data being captured outside of clinical studies and greater data sharing of clinical studies, external controls may become a more attractive alternative to randomized clinical trials. Both industry and regulators recognize that in situations where a randomized study cannot be performed, external controls can provide the needed contextualization to allow a better interpretation of studies without a randomized control. It is also agreed that external controls will not fully replace randomized clinical trials as the gold standard for formal proof of efficacy in drug development and the yardstick of clinical research. However, it remains unclear in which situations conclusions about efficacy and a positive benefit/risk can reliably be based on the use of an external control. This paper will provide an overview on types of external control, their applications and the different sources of bias their use may incur, and discuss potential mitigation steps. It will also give recommendations on how the use of external controls can be justified.

**1. Introduction: Short overview on what is going on from regulatory perspective**

The use of and interest in using Real World Data (RWD) within clinical development is increasing with the greater availability of a wide range of data sources, in particular with the widespread adoption of Electronic Health Records (EHR). The applications of RWD are widespread and include support of health technology assessments as well as design of clinical trials. One application that is gaining more and more visibility is the generation and use of external controls arms from RWD and other sources with the vision that these could fully or partially replace control arms in randomized clinical trials allowing trials to be run where a randomised trial would be unethical or impractical or smaller trials to facilitate quicker access to novel therapies for patients [8, 9, 10, 12, 13]. This could be specifically important for diseases with high unmet medical need. Other sources or external controls apart from RWD can be used using the same methodology, in particular the approach of using control arms from other clinical studies has a number of attractive features.

The Randomized Clinical Trial (RCT) is the established gold standard for medical research based upon its reduction of spurious causality and bias. RCT contribute to the generation of unbiased treatment effect estimates and are one pre-requisite for type I error control in hypothesis testing. Over recent years there has been an increasing number of single arm studies, particularly in the areas of oncology with phase 1 expansion cohorts, and in rare diseases. In addition we see an increasing number of studies where external controls were also used to supplement randomized clinical trials. Such studies have been used for both internal decision making and accelerated regulatory approvals [28, 29].

The use of real world data (RWD) or other sources of data, such as control arms from previous studies, as external control for a new study in clinical development is currently discussed by health authorities and industry to define its place in development, especially with regard to regulatory decision making on drug approval, see also the International Council for Harmonisation of Technical Requirements for Pharmaceuticals for Human Use (ICH) guideline on the choice of control group [11]. RWD have been used for a long time for the evaluation of long term safety data, even as a standard for post approval commitments. Its use as an external control arm in a non-randomized trial in evaluating efficacy is on the agenda today, and there is a willingness to consider such approaches but firm guidance or acceptance is not yet available. There are examples for both external control data being accepted and rejected for the primary analysis. Examples include the use of Real World Evidence (RWE) to Support Approval of Ibrance (Pfizer) for male breast cancer, the use of RWE to Support NDA for Selinexor (Karyopharm) for relapsed refractory multiple myeloma, RWD as an external control arm in Roche's entrectinib ROS1 positive metastatic non-small cell lung cancer submission, the approval of fish oil triglycerides, Omegaven (Fresenius Kabi), which was based on two single arm trials matched to a historical control arm from hospital records and Brinavess (Correvio). All these examples have in common that the decisions made were very situation-specific not permitting general guidance to be derived from them. Therefore, discussions between stakeholders on the usage of external control data are ongoing and will continue for the foreseeable future. There is an increasing hope of stakeholders that such single arm studies with external controls could speed up drug development and provide a cost effective option with limited and well understood impact on validity and generalizability of study results. Statisticians in companies as well as in regulatory and health technology assessment agencies should endeavor to ensure that they are fully involved in such discussions. Bias through intercurrent events can occur in RCTs after randomization as well as in external control studies, therefore RCTs are also not free of this bias. Sometimes, statisticians can insist on RCTs as the gold standard without well-reasoned argumentations with the risk of being cornered and not involved in further discussions. It will be important for statisticians to engage in these discussions and their expertise will help to find the best application of designs with external controls as a basis for decision making. When RWD is the potential source for the external control data, cross collaboration with epidemiologists and real world data experts will be key.

The view of health authorities on the use of external controls is still evolving and they have indicated that they are open to a dialogue on this topic. It can be expected that health authorities will maintain their current high standard of evidence for approval in the interest of public health and will need to be convinced that this is not being eroded as a result of the use of external controls. Maintaining the high standard for regulatory approval is also in the interest of companies with a strong interest in developing innovative medicine. Statisticians will play a key role in finding the right place for study designs using external controls and in defining how much the use of such designs could be expanded without substantially changing the high standard of evidence required for approval. As a result, we may discover that such designs may be used more frequently. The hope is that we can increase our insight in the risks and opportunities of external control studies to allow a better informed decision making on the best study design.

In clinical research, there are a variety of study designs, ranging from the gold standard of a RCT to a single arm trial with only external control data. Additionally, a number of designs use a combination of randomized control data and external control data, an approach which was discussed by Pocock in 1976 [44]. The randomized comparison of the test arm with just the patients randomised to control provides an unbiased estimate of treatment effect but may have higher variability and limited power, and this is always available as a fallback if the results from the randomised and external control arms differ, although this is likely to be an underpowered comparison. The external controls can be used to supplement the randomised controls, reducing the variability and giving a fully powered study. However, this comes at the cost of an increased risk of bias.

There are two main approaches for analyzing hybrid designs: static borrowing in which the external control data will be used irrespective of how the external data compares with the randomized control data, and dynamic borrowing designs, in which external controls are only used if they are consistent with the randomized control data for example by using a power prior approach [45]. Static borrowing carries the same risk of systematic bias as a single arm trial with external controls and may be difficult to interpret if the results from the two control arms differ. Dynamic borrowing designs, in contrast, can reduce the bias, but a residual bias still remains ([1].

 Single arm studies without any reference to external control data are simple and straightforward to analyze but do not allow any causality conclusions. Nevertheless, single arm studies are often used for causality assessments by comparing the achieved results with those reported in literature. Beside their simplicity, such designs are associated with a high risk of bias as discussed many times in the literature [6]. Where there is a substantial treatment effect and/or the level of response in the historical control arm is negligible we may feel confident that the observed treatment effect exceeds any biases and cannot be the consequence of undetected and unintended selection. For regulatory bodies it may be sufficient if the effect observed in a single arm study reliably exceeds all clinical experience in the patient population under investigation. However for HTA, a different design allowing establishing relative effectiveness would be preferable.

Threshold crossing is a simple way to utilize external control information without a direct comparison. Prior to observing any of the study data a threshold is defined based on external control data which needs to be exceeded by the active treatment arm in the single arm study. When this threshold is sufficiently conservative given external control data, a threshold crossing analysis allows transparent decision making on the efficacy versus external control. The most common examples of using threshold crossing are single arm studies in oncology where the threshold is often set to 20% overall response under the recognition that any response is highly unlikely in the absence of treatment and under the assumption that clinical benefit can be concluded from this.

This paper provides an overview on the use of studies with external controls, on their benefits and associated risks of bias. It provides an outline of standards, encourages statisticians as well as associated disciplines and their practitioners, e.g., epidemiologists and RWD/RWE experts in companies and health authorities to work together and outlines what needs to be done in order to allow more informed decision making on such designs in future. Section 2 will outline the importance of statisticians being engaged in these designs and section 3 will describe different sources of bias including potential mitigation steps. Section 4 will then discuss the need for more research to evaluate how the results of studies using external control arms could be further adjusted for residual bias and variability. Benefits of external controls will be described in section 5 and the need for pre-specification in section 6. Some examples will be given in section 7 and the paper will end with some conclusions.

## 2. Why is it important for statisticians to get engaged?

External controls are often promoted as having the potential to provide benefit to patients by speeding up access to new therapies and exposing less patients to a potentially suboptimal treatment as well as benefitting the company by potentially speeding up drug development. On the other hand, external control studies open up the risk for biased estimation of treatment effects and may increase the risk of false positive or false negative results. Health authorities and health technology assessment (HTA) bodies may trade potentially faster market access against risk on the long run for those who use the medication. A critical factor in the successful adoption of these new opportunities will be to maintain the same high level of scientific rigor associated with randomized clinical trials and to ensure that the quality requirements for regulatory decision making is not negatively impacted. Statisticians have a unique set of skills and understanding of adequate methodology which could help to define the best place for such new opportunities as well as ensuring external control data has a sufficiently high level of quality. It is therefore important that statisticians are fully engaged in the use of external controls and supporting clinical teams to find the right applications of external data for the benefit of patients.

## 3. Sources of bias when using external controls

The use of external controls unavoidably comes with the risk of biased estimates and hypothesis tests with inflated type 1 error rates. When using external control as comparator it is therefore important to address and discuss all sources of bias as we need to be aware of potential bias in the interpretation of results. Discussion and elimination of bias to the extent possible is therefore important even when we know that residual bias can never be completely ruled out. The discussion of biases should be done in a transparent way addressing also clearly the disadvantages and risks associated with external controls. The decision for or against a study with external control will depend on the size of anticipated biases as well as the possibility to adjust for them. The following is a summary of the most important types of bias associated with external controls, for a summary see also Table 1.

### 3.1 Selection bias

Patients enrolled in clinical trials can be different to patients being enrolled in previous trials or different to patients treated in clinical practice [18, 19, 20, 21, 22]. Patients in one trial may be different to patients in another trial, because of different institutions or regions being involved, different study criteria or many other reasons. Differences between study populations and RWD may come from institutions active in clinical research being different from other institutions or that patients enrolled in clinical trials frequently have better outcomes than patients not enrolled into trials. Some differences are explicit and intentional, i.e. through eligibility criteria such as legal age limits or laboratory value thresholds. Other differences are implicit and typically not intentional, e.g. sponsors preferring large or academic trial centers which limits the inclusion of the rural population or investigators hesitating to include frail patients into a trial that places additional burden on them. Consideration should be given to the impact of moving away from a randomised trial to a single arm trial with external control, as this may access additional patients that would not have enrolled into the randomised trial because of the possibility of receiving a placebo or treatment with a low likelihood of a positive outcome, and this expanded population may exhibit a different response to treatment.

The experiences of patients enrolled into clinical trials can also differ from patients in regular clinical practice, for example in the number and regularity of treatment visits, and the amount of attention received. Further differences also exist in evaluation criteria to follow pathology evolution as well as patient characterization. Additionally, registries and electronic health records (EHR) also have inherent biases important to consider ([37-39]).

Naïve unadjusted comparison of populations from different sources would be expected to lead to biases in any treatment comparison due to confounding with differences between the populations. Therefore, as a first step it is important to prospectively set clinical trial eligibility criteria based on measures collected in routine clinical practice and apply these criteria to the external control database. Afterwards, there is still likely to be residual differences between the populations, and techniques such as propensity scoring [42], matching and covariate adjustment will reduce further biases. But in the absence of randomisation, there is always the potential for unobserved confounding variables which will create a residual bias in any treatment comparison, and it will never be possible to completely eliminate the possibility of the presence of unobserved confounders [30, 31,32].

### *3.2 Calendar time bias*

Patients treated in the past may respond differently than those treated today due to changes in standard of care. Usually such calendar time bias will favor the read out of patients treated today, i.e. the experimental treatment group. One mitigation step is to use concurrent control patients from a real world data source. Alternatively, using non-concurrent patients may require explanation of why major changes in outcome from the standard of care are unlikely and therefore calendar time bias would be small. Examination of clinical trial results or real world data over a period of time may indicate the level of calendar time bias for an indication, but attention should be given to the possibility that these may be indication and region specific [33].

### *3.3 Regional bias*

Patient outcome may vary between regions due to differences in many factors, e.g., ethnicity, patient compliance or the health care system and this bias could go in either direction. The easiest way to avoid such bias is to use control patients from the same regions as the clinical study. If this is not feasible, one could argue that regional bias should be limited where there is evidence for limited differences in the level of health care between regions. Workups with randomized controlled studies, replacing the randomized control by external control patients from a different region may help to identify the size of regional bias and to help understanding in which scenarios the bias should be rather small. For regulatory agencies, it may be important that the external control population matches the actual trial population. For HTA bodies this may not be sufficient as the review of such bodies has a greater focus on subgroups. Finally, one could make a similar argument for comparability of centers or type of centers (in the same way as for regions, see also [23]).

### 3.4 Assessment bias

Knowledge of therapy can influence the outcome assessment. This assessment bias can be especially large for endpoints with a larger degree of subjectivity in their assessment. For endpoints with high subjectivity in their assessment, blinding in a RCT is possible for assessment bias. Blinding however is often not feasible with external controls. Choosing an objectively assessed endpoint, for example overall survival can drastically reduce assessment bias. Independent blinded review approaches for external controls data could be tried, but may have limited applicability. Choosing an endpoint with highly subjective assessment could limit the applicability of external controls.

### 3.5 Different endpoint bias

Endpoints in clinical trials may be different than endpoints in routine clinical practice or may not be even assessed in routine clinical practice, for example response and progression endpoints in clinical trials in oncology. In addition, they may be measured with a different level of quality. This may limit the applicability of external controls when using these endpoints [43]. Every effort should be made to obtain the same level and quality of information from external control data as in the study, for example by obtaining consent for tumor images to allow RECIST-like review. On top of this, the frequency of assessments in external control data may differ from studies and could introduce bias. In this case, we should conduct sensitivity analyses, e.g. to investigate the impact of the frequency of assessments on the outcome. We need to accept that this may not work for all EHR databases and could limit the application of external control data. Alternatively, one could also design a more life-like single arm study so that its design already matches available endpoints in external control databases. This may make sense in situations in which a single arm trial is the only viable option, even when the trial is then larger.

### 3.6 Immortal bias

Defining the study start or day 1 for a patient is necessary to provide an anchor to define the patient specific time scale which all analyses subsequently depend upon. In a study day 1 could be the date of randomization, the date of informed consent or the date of first dose, but can always be clearly identified. For external control data, especially when coming from registries or EHR, it is not always clear how to define day 1 for a patient. Differences could generate immortal bias, for example in a trial with an overall survival endpoint only including patients who due to the design of the data collection we know to have survived a period of time, which could be serious and limit the applicability of external control data. We therefore need to clearly define this time origin. Sensitivity analyses should be used to perform an evaluation of the potential bias introduced in cases where there is no obvious choice. And we should note that dealing with immortal bias could be very tricky and may require creative thinking, and that in some situations statistical solutions may not be obvious or may not even exist at all.

### 3.7 Retrospective selection bias

Bias resulting from retrospective selection can be serious, especially when selecting external data and key analysis features, when the external control results are already known. Pre-specification is therefore an essential pre-requisite when using external control data. Using concurrent controls would further help to eliminate retrospective selection bias or the perception of retrospective bias, as well as calendar bias. In rare diseases, the use of all available source of available patient level data as external controls may be another way to limit retrospective selection bias. The issue may be particularly serious in registries when data are often not complete and the selection of patients may be perceived as a cherry picking effort. Quite often assessing the nature and degree of selection bias going on can be difficult. Access to performing full assessments may be hampered by conflicting interests of registry holders and privacy right concerns, and in addition, it requires special technical expertise that is not necessarily readily available. Writing a prospective detailed protocol is critical, and although pre-specification is key, it does not solve all problems. For example, centers with a low historic response rate under the standard of care are likely to continue to have a low response rate in concurrent external control patients.

### *3.8 Study bias*

Patients in clinical trials have different outcomes compared to patients in clinical practice, often referred to as a so-called second translational gap [24, 25, 26] for numerous reasons including placebo effects and different level of care and compliance for patients in clinical studies. In addition, the concomitant treatment setting may not be controlled for external controls as can be done in a prospective trial. In order to mitigate this bias we can choose external control patients from previous similar clinical trials or even trial hospitals. However, this may not always work, especially when the external control patients come from studies with other sources of inherent bias such as calendar time or regional bias. Investigations of randomized controls studies, replacing randomized control patients with control patients from another study may help to evaluate the size and impact of this bias in a given indication. An alternative solution could be to apply platform trials [34, 35].

### *3.9 Between study variability*

An important issue is also how variable and heterogeneous the outcome is across different studies. The reliance on external controls will be more critical if there is a high between study variability that cannot be explained, as this indicates that important factors affecting the outcome of an endpoint in a study cannot be controlled for. In such a situation, an RCT may be strongly recommended. Examples could be studies in schizophrenia or depression where high placebo responses are often observed and where the biological mechanism of the disease may be less well understood making it extremely challenging to control for the resulting variability.

### *3.10 Intercurrent event bias after study entry*

Intercurrent events such as premature treatment discontinuation can occur in RCTs and in external control studies alike. Depending on the estimand intercurrent events can lead to substantial bias. The ICH E9 addendum [2] makes a strong effort to control for such bias in a transparent manner, and the ICH E9 addendum should be similarly applied for external control studies. In addition, the quality of external control data needs to be looked at, especially for a different frequency of intercurrent events or even different types of intercurrent events. Such differences can further impact the usability of a data source as external control and needs to be further investigated.

### *3.11 Some examples*

The types of bias encountered in using an external control is largely determined from its source. Selection bias can occur nearly with every source of data. Assessment and different endpoint bias may be more likely when using real world data than when using data from a previous trial. Calendar time bias may not be present when using concurrent data from another trial or from an EHR source. In general, RWD data from EHR are more prone to many sources of bias including differences in missing data, assessment bias and study bias than data from other sources. Also note, the advantage generally associated with RWD of reflecting wider clinical practice rather than the more controlled environment within a clinical trial is not wanted when using RWD as an external control for a study population. When using RWD as an external control patients need to be carefully selected so that the resulting control no longer reflects the real world but reflects the study population.

## 4. Adjustment for additional variability and bias

External controls generate additional variability and bias, the latter leading to alpha inflation of unknown size. Researchers have started to look into completed RCTs in a meta-analytic fashion, replacing the randomized control arm by an external control arm to see how well the arms match [8]. This is sometimes termed validation by replication, although it is unclear what the term validation means in this context. Such work is insufficient as a validation step in the sense that it would sufficiently ensure that external controls always work without substantial inflation of bias and variability, it may however be important as a proof of concept for external controls evaluating how good or bad external controls can be. These exercises if being more meaningful need to carefully justify the selection of studies and attempt to be as complete as possible in the inclusion of trials to avoid any selection bias on a study level.

These approaches may also form the basis for incorporating additional bias and variability into external control analyses. Methods need to be developed to take the additional variability and bias identified from sets of historical studies into account and incorporate these insights into new studies and into decision making. These methods may help in getting external control studies accepted in a broader set of applications and in situations where bias mitigation as discussed in section 3 is limited. The framework for applying studies with surrogate endpoints is a good example [27] which one may want to mimic here as well. The meta-regression framework for surrogate endpoints has allowed decisions when surrogate endpoints are acceptably close to clinical endpoints and on how to adjust for the additional uncertainty. As with surrogate endpoint validation there are some limitations: validation is likely only to be valid for one specific treatment or class of treatments and extrapolation beyond this may be challenging and such a framework for external control may be bound to a specific indication, endpoint and source of external control. More consideration needs to be given to understanding when meta-analysis approaches of existing studies can be extrapolated to a new study with external controls, how many studies are required to robustly generate such a relationship and how well relationships in a particular indication or class of drugs can be extrapolated to other scenarios.

## 5. Benefits and risks of external controls

One of the justifications given for adopting external controls is a reduction in time and cost, for the benefit of patients as well as for the company. However, the savings of these benefits may sometimes be overestimated. Reducing the sample size of a study by 50% does not reduce time and costs by 50% as well.

In addition, as mentioned earlier, increasing the sample size in a single arm study to allow for a larger than preferred heterogeneity in the trial population might make external controls more acceptable to regulators and HTAs. The additional heterogeneity can alleviate some of the concerns as this might make the trial more life-like and hence a better match to the external data source. Recruitment generally increases throughout the course of a trial and so adding patients at the end of the trial may not be that time consuming. Also a 400 patient trial compared to a 200 patient trial generally uses more centers, indicating again that the time gain may not be that much. Although the time needed to recruit patients is reduced, the time needed for planning and analyzing the trial are not reduced and given the increased complexity of a trial with external control these times are potentially longer than for a traditional RCT. For a time-to-event endpoint where the sample size is dependent upon the number of events with an efficacious new treatment, these events may occur very slowly, and may become the rate limiting step rather than observing enough events on the control arm. In the overall cost of a study only a part of the overall cost is attributable to patients. So, a 200 patient study may not be that much cheaper than a 400 patient study. It also worth considering that access to and the processing of RWD may generate additional costs.

This assessment of potential savings is primarily for studies with only an external control arm; for studies with a hybrid design using both, internal randomized and external control, the savings are rather smaller as the core clinical trial still includes randomized control patients. In studies where an external control is only used for long-term comparisons in chronic diseases, there are no savings in cost or time, the benefits come from an increase in the quality of conclusions compared to having no formal long-term comparison.

Any benefits of reducing study duration should be considered holistically of the overall time it will take to get a treatment to patients. There is little value in completing a study earlier if there are subsequent greater delays, for example due to either the regulatory or HTA approval process because of concerns about the interpretability of the results.

Another justification for adopting external controls is that the studies would be more attractive to patients by increasing enrollment rates. This may in fact be often true. However, this does call into question the assumptions of equipoise underlying the clinical trial and it would certainly be unethical to conduct a study unlikely to deliver interpretable conclusions or even deliver misleading conclusions.

One of the strongest arguments for the consideration of external controls lies in the patient's perspective and its potential ethical implications. We sometimes underestimate what randomization could mean for patients. Uncertainty around the treatment group being assigned or worries about suboptimal treatment when assigned to control could provide additional burden for the patients. A lot of patients participate in a trial for hope in a hopeless medical situation, especially in diseases like oncology. Randomization is not the right answer for them, and this is also why a randomised study is sometimes difficult to run. Studies in children are examples where the ethics around randomisation are often difficult. Further examples can be found with rare diseases. Regulators often recognize settings like these and accept single arm studies. Even when regulators primarily evaluate such single arm studies on their own merits, external control comparisons add substantial further evidence to the results. In other slowly progressing diseases, especially those requiring chronic lifelong treatments like in ophthalmology, randomisation could be maintained for one or two years but then patients demand to enter open label extension therapies. Short duration trials in such a disease are not informative and are also scientifically questionable as it is the long term outcome which counts most. In such situations, single arm trials or randomized trials with external controls for long term follow up could increase acceptance of the trial for current patients.

Ethical considerations must consider the ethics associated with patients entering a trial alongside the ethics of future patients and their demands for efficacious therapies. Future patients may have a strong interest in the increased strength of evidence associated with randomized experiments compared to external control studies. The risk of increased false positive findings in studies with external control and the possible consequences for patients receiving ineffective treatments has to be weighed against any concerns of study patients about randomization. Even when we need to balance both sides, the ethical argument or patient's perspective against randomization remains a valid and important one in some situations.

Finally, there is also a risk for industry which is often underestimated when embarking on single-arm clinical trials. Lack of fit of external control in the current treatment context and resulting bias could lead to wrong conclusions at the end, to false investment or a delay of a whole development program for years. This risk is not fully covered by a dynamic borrowing design as the consequence of a population mismatch would be an underpowered trial. The decision to embark on a single arm trial has therefore to be well thought through by considering benefits and risks.

## 6. The importance of quality

### 6.1 Quality of data

The outcome of any study can only be as good as the quality of its underlying data. This general principle is even more true for external controls, especially when RWD are used as external control and the quality of data is beyond the control of the investigator. When using other clinical trial data often the quality of the underlying control data then should be similar to the data from the active study. In any case, before external data can be used, a substantial effort has to be made to make them as complete as possible to a level which is comparable to the quality in the actual study. Typically, the external control data will have been collected for purposes other than medical research, e.g. medical records to document the provision of health care or claims data to make financial transactions. Different treatments and procedures by health care insurance or "upcoding" need to be addressed here. In addition, unlike for a clinical trial, contradictory information, e.g. when the concomitant treatment does not match the concomitant disease, cannot be queried and must be dealt with as it is. Finally, as already mentioned the amount of missing data and its implications for intercurrent events and estimands may further complicate the discussion around bias, see also section 3.4 and the ICH E9 addendum [2].

### 6.2 Quality of how we deal with sources of bias

RCTs may be more difficult or elaborate to perform (although this may not be true in many cases) and may require more patients but undoubtedly, they are much easier to interpret and their conclusions carry utmost weight. While studies with external control may be easier to perform and might be ethically more appealing, their main weakness lies in the strength of conclusions which could be made. Therefore, additional efforts are needed in the interpretation of a study with an external control, including a thorough consideration of underlying sources of bias and their likely impact on conclusions. It will be possible to address some sources of bias by design and/or by analysis. For remaining sources of bias, we should be transparent and acknowledge their existence and likely size and impact on conclusions. Sensitivity analyses investigating different assumptions on the size and nature of different biases may be valuable tools in interpretation.

### 6.3 Transparency

Similar to an open label trial, and probably even more so, there is a critical need for transparency on the processes and decision making on design features in a trial with external controls. The number of options to "game the system", by choosing design features up front that will have a "positive" impact on outcome in such a trial are larger than in a randomized clinical trial. Transparency is therefore key to build trust and avoid any perception of bias as much as possible. Pre-specification of all analysis steps is of crucial importance in this context, together with the ability to demonstrate that key features of the design and analysis strategy have actually been truly pre-specified. Transparency should include also information on how external control data, especially RWD data were handled as such data sources may not be that easy to assess upfront in terms of quality and relevance.

In addition, patient level data of most clinical trials can be freely accessed by independent researchers as all major pharmaceutical companies have pledged to share data [14]. However, RWD are typically not freely available and independent researchers might find it difficult if not impossible to reanalyze the data of a trial with external control.

## 7. Examples and applications

Decision making on when to use an external control study should for the foreseeable future be primarily based on ethical considerations or feasibility. RCTs are likely to remain the gold standard for the foreseeable future, and so moving away from the RCT requires strong valid reasons. As discussed in the previous section, arguments for cost and time could be overestimated and should be secondary to ethical concerns, although cost and time savings do have ethical dimensions in terms of releasing funding for developing other therapies or making new therapies available earlier.

Nevertheless, the appropriateness of when external control arms should be considered will largely be scenario-dependent. For an experimental treatment with a novel mechanism of action in a high prevalence disease with well-established and effective standard of care, it would be difficult to propose anything other than a RCT. The application of external control arms may be more acceptable in settings with some or all of the following properties: a rare disease, no effective standard of care available, a robust endpoint, large game-changing treatment benefits or a molecule with efficacy already established in other diseases or indications.

For any study which could involve external data sources, any precedence or understanding of the attitude of the regulatory and/or payer authorities to external controls arms within that setting should be considered. Finally, marketing and pricing considerations should be factored in the decision making as well as for the foreseeable future RCTs will be viewed as providing the highest level of evidence and therefore count more for payers and key opinion leaders than studies using external controls. It is important to engage in upfront discussions with regulators and HTA bodies on the level of evidence provided by such studies and use a design which satisfies all stakeholders.

### 7.1 Traditional: Designs using threshold crossing (e.g. oncology, contraceptives)

Examples here include single arm trials in end-stage diseases in oncology where no accepted therapy exists and with response rate as the primary endpoint, see for example [36]. In these situations, there is an ethical demand not to randomize versus best supportive care as being randomised to supportive care leaves patients with no hope of an efficacious treatment. The underlying design is typically a threshold crossing design [13] assuming that with best supportive care no tumor response would be observed. Usually the required level of response is not defined by external control as the assumption is no response with best supportive care. The argument is that a compelling response rate under the new therapy would be sufficient. Adjustments for other endpoints beyond overall response are usually not made although external controls could provide a valid comparison and would actually increase the overall quality of the study. There are also scenarios where there is an accepted standard of care therapy, but with limited efficacy.

Threshold crossing designs are typical for oncology phase II studies in an end stage disease, where patients do not have any treatment options left. Threshold crossing can however also be used in other indications outside oncology where we have a strong endpoint which can reflect strong activity sufficiently differentiating from standard of care. Patients with spinal muscular atrophy (SMA) of type 1 are an example. This disease starts quickly after birth, and babies with SMA type 1 usually do not survive two years and never sit by themselves in their lives. A new treatment can then be tested for example with an endpoint "unassisted sitting for at least 5 seconds" or "one year survival" in a single arm trial when the threshold is suitably selected, the study provides evidence to differentiate from this threshold [46].

Another example where single arm trials are accepted are contraceptives where the pregnancy rate should be less than 1% compared to some 80% of women getting pregnant per year without birth control [15], section 3.5. Only in case of new products comparing against a method with an expected pregnancy rate above 1% (e.g. condoms or fertility tracing apps) a comparative trial is required.

### 7.2 Situations where randomized control is not feasible

This scenario includes situations in rare diseases where we often see limitations of being able to recruit a sufficiently large number of patients into the trial. Examples include trials in male breast cancer patients, and randomized trials in children which come with specific ethical burdens, especially in rare diseases. For pediatric patients with a serious, life threatening disease, recruitment into randomized trials is often challenging as randomization can be unacceptable to parents, especially when the outcome without active therapy is almost certainly negative. Another class of studies where randomized control is difficult are extension studies as these usually start as randomized clinical trials where randomization can be maintained only for a limited time period and patients are subsequently switched into an open label extension where important clinical information is gathered without adequate control. External controls could make medically relevant long term follow up information more readily interpretable and useful. In these situations, health authorities have generally been open to external control trials, HTA bodies generally less so. It is therefore essential to seek guidance from all relevant health authorities and HTA bodies before starting such a study. A challenge in this situation, especially for rare diseases, is that sometimes only scarce external data is available which may be of insufficient quality or may come with a high risk of selection bias. Such problems might hamper the use of such comparisons, irrespective of the willingness of regulators and HTAs to accept such approaches.

### 7.3 Designs with both randomized and external control

There are further designs beyond hybrid designs with a mix between randomized and external controls. Long term extension studies are one example where the randomized period could be used to assess the quality of external controls and the potential level of bias in the long term.

Other study designs could provide fall back positions, for example adaptive designs where the first stage includes both randomized and external controls followed by an interim analysis which decides whether the second stage will use a randomized control or an external control. It is always important to keep alert to opportunities and to use the best possible design to speed up decision making without taking more risks or substantially changing the standards of regulatory decision making. All these alternative study designs require a rigorous discussion on the degree to which they still allow robust conclusions on causality.

Another application are platform trials, where multiple treatments are compared to a control and treatment arms may enter or leave the trial over time such that the active treatment arms vary over the study period. For such trials it has been proposed to base treatment control comparisons not only on control group data collected in study periods where the respective treatment is part of the platform, but also non-concurrent controls from other time periods. While these non-concurrent controls are not entirely "external" since they are recruited in the same platform trial, they can be subject to similar biases as external controls [35]. Especially, time trends in the control group data can bias the analysis and several approaches to adjust for this bias have been proposed [40, 41].

### 7.4 External control studies in early development

Another example of using external control studies is in early development purely for internal decision making where the risk lies solely with the sponsor. These studies typically have small sample sizes, higher rates of risk and often use different endpoints compared to later stage studies. Whilst the decisions made from these studies are at the company's risk, it is still important for the company to be able to accurately understand the level of risk associated with any potential go or no-go decision, otherwise, reliable portfolio decisions are not feasible. The company should give the same level of consideration to the biases and risks inherent in these designs as to later phase designs.

Such early trials could be used as a testing ground to gain more experience with external controls, and this experience may result in improvements in the quality of external data sources. Nevertheless, good decision making is equally important in early as in late stage studies and therefore similar principles should be applied as for late stage studies. It may however be easier to experiment and opt for an external control design in early development, as there are fewer stakeholders than in late stage. There are examples in early development where external controls make sense and improve the quality of decision making, for example using multistate models in oncology [7]. More systematic reporting and sharing of the impact of using external data on industry decisions could further strengthen the credibility of studies with external controls more generally.

### 7.5 The role of frequentist hypothesis testing in studies with external controls

The role of hypothesis tests and p-values may change in studies with external control and their interpretation may be more difficult. When using the threshold crossing approach, for example, the hypothesis test of a single arm study shows that the new therapy statistically separates from the chosen threshold (testing the hypotheses H0: Response ≤ threshold versus H1: Response > threshold). While this is a valid test for this hypothesis, a meaningful interpretation as demonstration of a treatment effect will critically depend on the choice of the threshold. To account for sampling variation in the control group, the threshold could initially be chosen as 97.5% confidence bound for the control, but to robustly demonstrate a treatment effect, this threshold will typically have to be adjusted for other biases ([6]). Then the hypothesis test of the single arm trial is also a valid (but typically strictly conservative) test of the null hypothesis of no difference between the treatment and external control group (see e.g. [47]). The biases discussed in section 3 indicate the presence of confounders which could also drive differences in results, beyond those attributable to treatment effects. Further adjustment of the threshold for selection bias and adequate control of other bias may leave hypothesis tests sufficiently robust. Nevertheless, as the quantification of these biases is difficult, the interpretation of hypothesis tests in these cases remain challenging. Similar considerations hold if hypothesis tests directly comparing treatment and external controls are used.

While dynamic borrowing designs can limit bias and type I error inflation to some extent, they cannot fully control it. In fact, it has been shown that under very general circumstances, dynamic borrowing will not maintain type 1 error control [1]. Simulations show that for dynamic borrowing designs with a nominal type I error of 5%, the true type I error may be inflated up to 7-10% in the worst case. This inflation could in principle be corrected by lowering the nominal significance level to enforce strict type I error control, but this would result in any power gains from dynamic borrowing being completely lost [4, 5]. As a result, the window of applications for these designs may be narrow if strict type 1 error control is of importance, and for regulatory application. Upfront discussions with health authorities are therefore essential. More information can be found in the FDA guideline on adaptive designs [2] indicating that FDA is willing to consider a broader concept of controlling erroneous conclusions (i.e. a balancing between type-1 and type-II errors).

### 7.6 More general considerations: What do we need?

To achieve the goal of making drug development more efficient we should never forget that there are a broad range of options, tools and study designs available. Using external control studies is just one option, but in any particular situation it may not be the most effective choice to improve drug development. While external controls can improve decision making quality in certain situations and make drug development more efficient in others, we have to remind ourselves that there are also other methods which could also improve efficiency and that we have to be equally open to these alternatives, always applying the best and most efficient method. In addition, some methods fulfill the requirements on decision making of one stakeholder better than for others, and methods fulfilling all stakeholders' requirements will be preferred.

## 8. Conclusions

For the foreseeable future RCTs are likely to remain the gold standard for generating clinical evidence with external control trials used in situations where RCTs are not feasible or not ethical. In these situations it will be important that statisticians are engaged in designing these studies to optimize their value. Statisticians in industry, health authorities, health technology agencies, and academia should support the adoption and use of all new data sources, including RWD, which can assist in either improving decisions made within drug development or allowing drug development to proceed either more rapidly or in a way that better meets patients' needs. Statisticians should continue to acknowledge the RCT as the gold standard for robust data generation and conclusions on causality, but at the same time recognize that in some circumstances a RCT may be neither practical nor optimal from the overall perspective of a drug development program and so alternatives, for example utilizing external controls, should be considered. Some methodology for these approaches has already been developed and statisticians should continue to help to further refine these approaches. Any use of alternative data sources should go in hand with a rigorous understanding of any potential bias and variability which may impact the robustness of any decisions. It is also of key importance to ensure that the level of evidence required for any decisions, whether internal or external, remains at an appropriate level.

Statisticians should stay open to the option of single arm trials with external controls or hybrid designs wherever it makes sense. Yet, they should also refrain from jumping from the rejection of the RCT as the optimal design to the immediate acceptability of the external control as the preferred option. Instead, all design options should be carefully evaluated within the context of the planned study before concluding that an external control approach is the scientifically best acceptable design. If a study is to be conducted with external control, statistical leadership is essential to maintain rigor and quality and, in the interest of patients, regulators and companies developing innovative medicines, the level of evidence required for approval should be maintained. Statisticians should further help to identify the extent to which external control designs could be used in a meaningful way and should support such designs where it makes sense. We need to keep in mind that there are situations where designs with external control clearly make sense and equally examples where they do not. As the field is still evolving, careful assessment of each situation and its unique characteristics will be important to identify the best possible path forward. Statisticians will be important participants in such discussions, particularly with clinicians and pharmaco-epidemiologists and it will be important to engage in discussion with regulators and HTA bodies as early as possible.

When designing trials with external controls we need to focus on data quality, data identification and qualification, and on a rigorous discussion of underlying types of bias. Wherever bias can be eliminated, or at least substantially decreased, this will ease interpretation and increase the likelihood of acceptance. Transparency in the design is key and we should pre-specify as much of the data processing and analysis as possible.

We should be aware that a randomized clinical trial will always have higher levels of credibility in the eyes of stakeholders such as regulators and HTAs and the medical community, due to the full control of many important sources of bias. Where multiple therapies are targeting the same indication, therapies tested in randomised trials will have an advantage in the marketplace from being supported by higher quality data. Therefore, decisions should not only be based on savings of time or investment. Ethical considerations should be the most important factor when deciding for or against external control. Considerations should also adequately balance type I and type II error as there are situations like in rare diseases with a high unmet medical need, where we might accept a less restrictive control of type I error for a gain in type II error control.

Finally, studies with external controls are only one of many ways to introduce innovation into clinical development. Even though we do not expect them to replace the RCT as the gold standard in the foreseeable future, they are an important additional tool for statisticians to have in their toolbox, and in some situations they will be the option of choice and other situations they will not. Statisticians should of course not forget all the other options available to innovate drug development and always stay open to alternative innovative designs. External control designs are an opportunity to further involve statisticians in discussions in terms of benefits and limits of design features considering benefit/risk aspects while maintaining a high level of evidence and generalizability for the study. It is in the best interest for patients and the company to choose a design which optimizes the development path for a molecule in a given situation.

**Acknowledgement**

Martin Posch is a member of the EU Patient-centric clinical trial platform (EU- PEARL). EU-PEARL has received funding from the Innovative Medicines Initiative 2 Joint Undertaking under grant agreement No 853966. This Joint Undertaking receives support from the European Union's Horizon 2020 research and innovation programme and EFPIA and Children's Tumor Foundation, Global Alliance for TB Drug Development non-profit organisation, Springworks Therapeutics Inc. This publication reflects the author's views. Neither IMI nor the European Union, EFPIA, or any Associated Partners are responsible for any use that may be made of the information contained herein.

| Potential Bias | Mitigation Strategy |
| --- | --- |
| *Selection bias*<br><br>*Patients enrolled in clinical trials are different in some ways compared to patients treated in clinical practice.* | Prospectively set clinical trial inclusion/exclusion criteria (I/E) based on measures collected in routine clinical practice. Apply I/E to external control database. Adjust for remaining population differences using propensity scores methods and/or multivariate analyses. |
| *Calendar time bias*<br>*Patients treated in the past do differently than those treated today.* | Taking control patients from the same calendar time as clinical trial or having evidence that standard of care & how it is used in this setting did not change<br>Otherwise: Show data indicating acceptable variability |
| *Regional bias*<br>*Patient outcome may vary between regions.* | Taking patients from the same region or having evidence that standard of care treatment and outcomes does not vary between regions<br>Otherwise: Show data indicating acceptable variability |
| *Assessment bias*<br>*Knowledge of therapy can influence the outcome assessment.* | Choose objective endpoints, e.g. OS, independent review (similar to strategies for open-label randomized studies).<br><br>Conduct sensitivity analyses |
| *Different endpoint bias*<br>*Response and progression in clinical trials are measured differently than in routine clinical practice. (This bias not applicable to OS.)* | Obtain consent for tumor images to allow RECIST-like review.<br><br>Conduct sensitivity analyses, e.g. frequency of tumor assessments. |
| *Immortal bias*<br>*Study start is difficult to determine as an anchor for patient specific time scale* | Need to clearly define time origin. Evaluation of potential bias introduced |
| *Retrospective selection bias*<br>*Select retrospectively external data and key analysis features.* | Pre-specification |
| *Study bias*<br><br>*Patients in clinical trials have different outcomes than in clinical practice. (e.g. due to placebo effect, different care)* | Show data indicating acceptable variability on outcome |
| *Between study variability*<br><br>*Reliance on external controls will be more critical if there is a high between study variability that cannot be explained, as this indicates that important factors affecting the outcome of an endpoint in a study cannot be controlled for.* | In such a situation a RCT may be mandatory or at least strongly recommended. |
| *Intercurrent event bias after study entry*<br><br>*Intercurrent events after study entry may be correlated with outcome and jeopardize comparability with external control* | Same bias as for RCTs. Same methods (see ICH E9 addendum) should be applied. |

**Table 1: Overview on the different sources of bias**